# Fuzzy Logic Based Method for Improving Text Summarization

Ladda Suanmali[1], Naomie Salim[2] and Mohammed Salem Binwahlan[3]
[1]Faculty of Science and Technology, Suan Dusit Rajabhat University, Bangkok, Thailand 10300
[2,3]Faculty of Computer Science and Information System, Universiti Teknologi Malaysia 81310
E-mail: [1]ladda_sua@dusit.ac.th, [2] naomie@utm.my, [3] moham2007med@yahoo.com

*Abstract*—Text summarization can be classified into two approaches: extraction and abstraction. This paper focuses on extraction approach. The goal of text summarization based on extraction approach is sentence selection. One of the methods to obtain the suitable sentences is to assign some numerical measure of a sentence for the summary called sentence weighting and then select the best ones. The first step in summarization by extraction is the identification of important features. In our experiment, we used 125 test documents in DUC2002 data set. Each document is prepared by preprocessing process: sentence segmentation, tokenization, removing stop word, and word stemming. Then, we used 8 important features and calculate their score for each sentence. We proposed text summarization based on fuzzy logic to improve the quality of the summary created by the general statistic method. We compared our results with the baseline summarizer and Microsoft Word 2007 summarizers. The results show that the best average precision, recall, and f-measure for the summaries were obtained by fuzzy method.

## I. INTRODUCTION

An increasingly important task in the current era of information overload, text summarization has become an important and timely tool for helping and interpreting the large volumes of text available in documents.

The goal of text summarization is to present the most important information in a shorter version of the original text while keeping its main content and helps the user to quickly understand large volumes of information. Text summarization addresses both the problem of selecting the most important sections of text and the problem of generating coherent summaries. This process is significantly different from that of human based text summarization since human can capture and relate deep meanings and themes of text documents while automation of such a skill is very difficult to implement. Automatic text summarization researchers since Luhn work [1], they are trying to solve or at least relieve that problem by proposing techniques for generating summaries. The summaries serve as quick guide to interesting information, providing a short form for each document in the document set; reading summary makes decision about reading the whole document or not, it also serves as time saver. A number of researchers have proposed techniques for automatic text summarization which can be classified into two categories: extraction and abstraction. Extraction summary is a selection of sentences or phrases from the original text with the highest score and put it together to a new shorter text without changing the source text. Abstraction summary method uses linguistic methods to examine and interpret the text. Most of the current automated text summarization system use extraction method to produce summary. Automatic text summarization works best on well-structured documents, such as news, reports, articles and scientific papers.

The first step in summarization by extraction is the identification of important features such as sentence length, sentence location [11], term frequency [6], number of words occurring in title [5], number of proper nouns [14] and number of numerical data [13]. In our approach, we utilize a feature fusion technique to discover which features out of the available ones are most useful.

In this paper, we propose text summarization based on fuzzy logic method to extract important sentences as a summary. The rest of this paper is organized as follows. Section II presents the summarization approach. Section III describes preprocessing and the important features. Section IV and V describes our proposed, followed by experimental design, experimental results and evaluation. Finally, we conclude and suggest future work that can be carried out in Section VI.

## II. SUMMARIZATION APPROACHES

In early classic summarization system, the important summaries were created according to the most frequent words in the text. Luhn created the first summarization system [1] in 1958. Rath et al. [2] in 1961 proposed empirical evidences for difficulties inherent in the notion of ideal summary. Both studies used thematic features such as term frequency, thus they are characterized by surface-level approaches. In the early 1960s, new approaches called entity-level approaches appeared; the first approach of this kind used syntactic analysis [3]. The location features were used in [4], where key phrases are used dealt with three additional components: pragmatic words (cue words, i.e., words would have positive or negative effect on the respective sentence weight like significant, key idea, or hardly); title and heading words; and structural indicators (sentence location, where the sentences appearing in initial or final of text unit are more significant to include in the summary.

In statistical method [14] was described by using a Bayesian classifier to compute the probability that a sentence in a source document should be included in a summary. [18] proposed a language- and domain-independent statistical-based method for single-document extractive summarization. They shown that maximal frequent sequences, as well as single words that are part of bigrams repeated more than once in the text, are good terms to describe documents.

In this paper, we propose important sentence extraction using fuzzy rules and fuzzy set for selecting sentences based on their features. Fuzzy logic techniques in the form of approximate reasoning provide decision-support and expert systems with powerful reasoning capabilities. The permissiveness of fuzziness in human thought processes suggests that much of the logic behind human reasoning is not only a traditional two-values or multi-valued logic, but also logic with fuzzy truths, fuzzy connectives, and fuzzy rules of inference [15]. Fuzzy set proposed by Zadeh [9] is a mathematical tool for dealing with uncertainty, imprecision, vagueness and ambiguity. Fuzzy logic in text summarization needs more investigation. A few studies were done in this area, Witte and Bergler [10] presented a fuzzy-theory based approach to co-reference resolution and its application to text summarization. Automatic determination of co-reference between noun phrases is fraught with uncertainty. Kiani and Akbarzadeh [12] proposed technique for summarizing text using combination of Genetic Algorithm (GA) and Genetic Programming (GP) to optimize rule sets and membership function of fuzzy systems.

The feature extraction techniques are used to obtain the important sentences in the text. For instance, Luhn [1] looked at the frequency of word distributions should imply the most important concepts of the document. Some of features are used in this research such as sentence length. Some sentences are short or some sentences are long. What is clear is that some of the attributes have more importance and some have less, so they should have balance weight in computations and we use fuzzy logic to solve this problem by defining the membership functions for each feature.

## III. EXTRACTION OF FEATURES

### A. Data set and preprocessing

We used 125 documents from DUC2002 to create automatic single document summarization. Each document consists of 8 to 60 sentences with an average of 28 sentences. The DUC2002 collection provided [8]. Each document in DUC2002 collection is supplied with a set of human-generation summaries provided by two different experts. While each expert was asked to generate summaries of different length, we use only generic 100-word variants.

Currently, input document are of plain text format. There are four main activities performed in this stage: Sentence Segmentation, Tokenization, Removing Stop Word, and Word Stemming. Sentence segmentation is boundary detection and separating source text into sentence. Tokenization is separating the input document into individual words. Next, Removing Stop Words, stop words are the words which appear frequently in document but provide less meaning in identifying the important content of the document such as 'a', 'an', 'the', etc.. The last step for preprocessing is Word Stemming; Word stemming is the process of removing prefixes and suffixes of each word.

### B. Sentence Features

After this preprocessing, each sentence of the document is represented by an attribute vector of features. These features are attributes that attempt to represent the data used for their task. We focus on eight features for each sentence. Each feature is given a value between '0' and '1'. There are eight features as follows:

*1) Title feature*

The word in sentence that also occurs in title gives high score. This is determined by counting the number of matches between the content words in a sentence and the words in the title. We calculate the score for this feature which is the ratio of the number of words in the sentence that occur in the title over the number of words in title.

$$S\_F1(S) = \frac{No.\text{Title word in } S}{No.\text{Word in Title}} \quad (1)$$

*2) Sentence Length*

This feature is useful to filter out short sentences such as datelines and author names commonly found in news articles. The short sentences are not expected to belong to the summary. We use the length of the sentence, which is the ratio of the number of words occurring in the sentence over the number of words occurring in the longest sentence of the document.

$$S\_F2(S) = \frac{No.\text{Word occurring in } S}{No.\text{Word occurring in longest sentence}} \quad (2)$$

*3) Term Weight*

The frequency of term occurrences within a document has often been used for calculating the importance of sentence. The score of a sentence can be calculated as the sum of the score of words in the sentence. The score of important score $w_i$ of word $i$ can be calculated by the traditional *tf.idf* method as follows [17]. We applied this method to *tf.isf* (Term frequency, Inverse sentence frequency).

$$w_i = tf_i \times isf_i = tf_i \times \log\frac{N}{n_i} \quad (3)$$

where $tf_i$ is the tern frequency of word $i$ in the document, $N$ is the total number of sentences, and $n_i$ is number of sentences in which word $i$ occurs. This feature can be calculated as follows.

$$S_{F3(S)} = \frac{\sum_{i=1}^{k} W_i(S)}{Max(\sum_{i=1}^{k} W_i(S_i^N))} \quad (4)$$

$k$ is number of words in sentence.

*4) Sentence Position*

Whether it is the first 5 sentences in the paragraph, sentence position in text gives the importance of the sentences. This feature can involve several items such as the position of a sentence in the document, section, and paragraph, etc., proposed the first sentence is highest ranking. The score for this feature: we consider the first 5 sentences in the paragraph. This feature score is calculated as the following equation (5).

$$S\_F4(S) = 5/5 \text{ for } 1^{st}, 4/5 \text{ for } 2^{nd}, 3/5 \text{ for } 3^{rd},$$
$$2/5 \text{ for } 4^{th}, 1/5 \text{ for } 5^{th},$$
$$0/5 \text{ for other sentences} \quad (5)$$

*5) Sentence to Sentence Similarity*

This feature is a similarity between sentences. For each sentence $S$, the similarity between $S$ and each other sentence is computed by the cosine similarity measure with a resulting value between 0 and 1 [20]. The term weight $w_i$ and $w_j$ of term t to n term in sentence $S_i$ and $S_j$ are represented as the vectors. The similarity of each sentence pair is calculated based on similarity formula (6).

$$Sim(S_i, S_j) = \frac{\sum_{t=1}^{n} w_{it} \times w_{jt}}{\sqrt{\sum_{t=1}^{n} w_{it}^2} \times \sqrt{\sum_{t=1}^{n} w_{jt}^2}} \quad (6)$$

The score of this feature for a sentence $S$ is obtained by computing the ratio of the summary of sentence similarity of sentence $S$ with each other sentence over the maximum of summary (7)

$$S_{FS(S)} = \frac{\sum Sim(S_i, S_j)}{Max(\sum Sim(S_i, S_j))}$$

*6) Proper Noun*

The sentence that contains more proper nouns (name entity) is an important and it is most probably included in the document summary. The score for this feature is calculated as the ratio of the number of proper nouns that occur in sentence over the sentence length.

$$S\_F6(S) = \frac{No.\ Proper\ nouns\ in\ S}{Sentence\ Length\ (S)} \quad (8)$$

*7) Thematic Word*

The number of thematic word in sentence, this feature is important because terms that occur frequently in a document are probably related to topic. The number of thematic words indicates the words with maximum possible relativity. We used the top 10 most frequent content word for consideration as thematic. The score for this feature is calculated as the ratio of the number of thematic words that occur in the sentence over the maximum summary of thematic words in the sentence.

$$S\_F7(S) = \frac{No.\ Thematic\ word\ in\ S}{Max(No.\ Thematic\ word)} \quad (9)$$

*8) Numerical Data*

The number of numerical data in sentence, sentence that contains numerical data is important and it is most probably included in the document summary. The score for this feature is calculated as the ratio of the number of numerical data that occur in sentence over the sentence length.

$$S\_F8(S) = \frac{No.\ Numerical\ data\ in\ S}{Sentence\ Length\ (S)} \quad (10)$$

IV. THE METHODS

The goal of text summarization based on extraction approach is sentence selection. One of the methods to obtain the suitable sentences is to assign some numerical measure of a sentence for the summary called sentence weighting and then select the best ones. Therefore, the features score of each sentence that we described in the previous section are used to obtain the significant sentences. In this section, we use two methods to extract the important sentences: text summarization based on general statistic method (GSM) and fuzzy logic method. The system consists of the following main steps:

(1) read the source document into the system;
(2) for preprocessing step, the system extracts the individual sentences of the original documents. Then, separate the input document into individual words. Next, remove stop words. The last step for preprocessing is word stemming;
(3) each sentence is associated with vector of eight features that described in Section III, whose values are derived from the content of the sentence;
(4) the features are calculated to obtain the sentence score base on general statistic method (GSM) shows in Figure 1. and fuzzy logic method shows in Figure 2.;
(5) a set of highest score sentences are extracted as document summary based on the compression rate.

*A. Text Summarization based on General Statistic Method (GSM)*

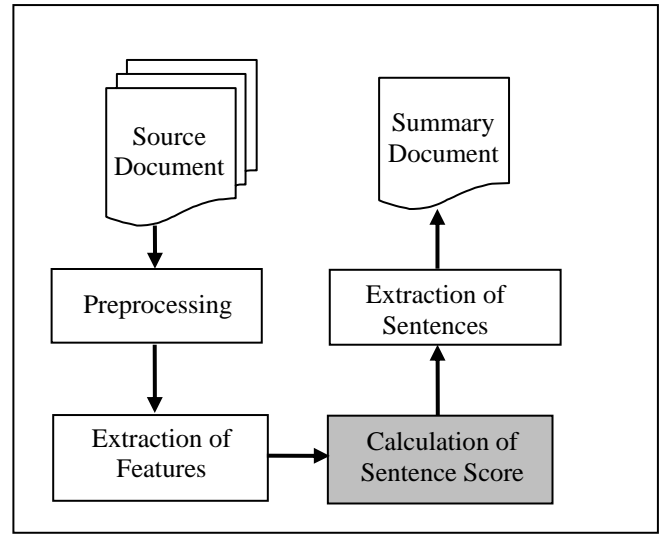

Figure 1. Text summarization based on general statistic method (GSM) architecture

Text summarization base on general statistic method is produced by the sentence weighting. First, for a sentence *s*, a weighted score function, as shown in the following equation is exploited to integrate all the eight feature scores mentioned in Section III

$$Score(S) = \sum_{k=1}^{8} S_{Fk(s)} \qquad (11)$$

*Score(S)* = The score of the sentence *S*
*S_Fk(S)* = The score of the feature *k*

### B. Text Summarization based on Fuzzy Logic

Fuzzy logic system design usually implicates selecting fuzzy rules and membership function. The selection of fuzzy rules and membership functions directly affect the performance of the fuzzy logic system.

The fuzzy logic system consists of four components: fuzzifier, inference engine, defuzzifier, and the fuzzy knowledge base. In the fuzzifier, crisp inputs are translated into linguistic values using a membership function to be used to the input linguistic variables. After fuzzification, the inference engine refers to the rule base containing fuzzy IF-THEN rules to derive the linguistic values. In the last step, the output linguistic variables from the inference are converted to the final crisp values by the defuzzifier using membership function for representing the final sentence score.

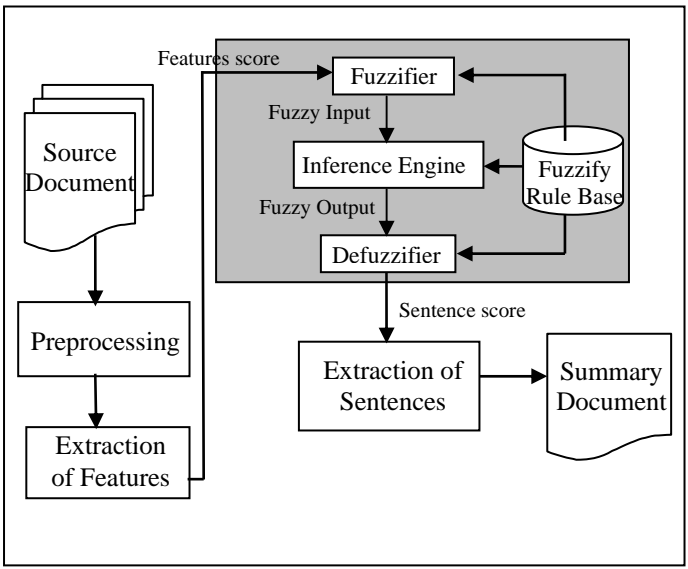

Figure 2. Text summarization based on fuzzy logic system architecture

In order to implement text summarization based on fuzzy logic, first, the eight features extracted in the previous section are used as input to the fuzzifier. We used Triangular membership functions and fuzzy logic to summarize the document. The input membership function for each feature is divided into five fuzzy set which are composed of unimportant values (low (L) and very low (VL), Median (M) and important values (high (H) and very high (VH)).

The generalized Triangular membership function depends on three parameters a, b, and c as given by (12) [19]. A value from zero to one is obtained for each sentence in the output based on sentence features and the available rules in the knowledge base. The obtained value in the output determines the degree of importance of the sentence in the final summary.

$$f(x, a, b, c) = \max\left(\min\left(\frac{x-a}{b-a}, \frac{c-x}{c-b}\right), 0\right) \qquad (12)$$

The parameters *a* and *c* set the left and right "feet" or base points, of the triangle. The parameter *b* sets the location of the triangle peak. For instance, membership function of number of words in sentence occurred in title is show in Figure 3.

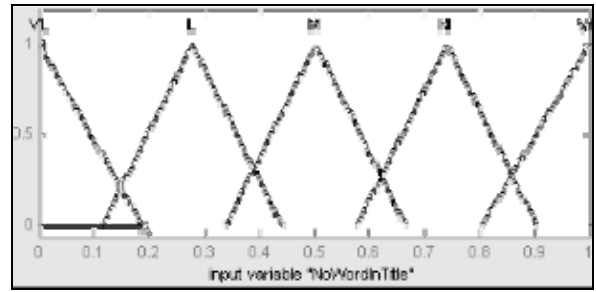

Figure 3. Membership function of number of words in sentence occurred in title

In inference engine, the most important part in this procedure is the definition of fuzzy IF-THEN rules. The important sentences are extracted from these rules according to our features criteria. Sample of IF-THEN rules shows as the following rule.

IF (NoWordInTitle is VH) and (SentenceLength is H) and (TermFreq is VH) and (SentencePosition is H) and (SentenceSimilarity is VH) and (NoProperNoun is H) and (NoThematicWord is VH) and (NumbericalData is H) THEN (Sentence is important)

Likewise, the last step in fuzzy logic system is the defuzzification. We used the output membership function which is divided into three membership functions: Output {Unimportant, Average, and Important} to convert the fuzzy results from the inference engine into a crisp output for the final score of each sentence.

### C. Extraction of Sentences

Both GSM and fuzzy logic method, each sentence of the document is represented by sentence score. Then all document sentences are ranked in a descending order according to their scores. A set of highest score sentences are extracted as document summary based on the compression rate. Therefore, we extracted the appropriate number of sentences according to 20% compression rate. It has been proven that the extraction of 20% of sentences from the source document can be as informative as the full text of a document [16]. Finally, the summary sentences are arranged in the original order.

## V. EVALUATION AND RESULTS

We use the ROUGE, a set of metrics called Recall-Oriented Understudy for Gisting Evaluation, evaluation toolkit [7] that has become standards of automatic evaluation of summaries. It compares the summaries generated by the program with the human-generated (gold standard) summaries [18]. For comparison, it uses n-gram statistics. Our evaluation was done using n-gram setting of ROUGE, which was found to have the highest correlation with human judgments at a confidence level of 95%. It is claimed that ROUGE-1 consistently correlates highly with human assessments and has high recall and precision significance test with manual evaluation results. We choose ROUGE-1 as the measurement of our experiment results. In the table 1, we compare the average precision, recall and f-measure score between general statistic method (GSM), fuzzy summarizer, Microsoft Word 2007 Summarizer and baseline summarizer form DUC2002 data set. The baseline is the first 100 words from the beginning of the document as determine by DUC 2002.

TABLE I THE COMPARISON AVERAGE PRECISION, RECALL AND F-MEASURE SCORE AMONG FOUR SUMMARIZERS

| Summarizer | Average | | |
|---|---|---|---|
| | *Precision* | *Recall* | *F-measure* |
| GSM | 0.49094 | 0.43565 | 0.45542 |
| Fuzzy | 0.49769 | 0.45706 | 0.47181 |
| MS-Word | 0.47242 | 0.40778 | 0.43026 |
| Baseline | 0.47002 | 0.45624 | 0.46108 |

The results are shown in Table I, GSM reaches the average precision of 0.49094, recall of 0.43565 and f-measure of 0.45542. The fuzzy summarizer achieves the average precision of 0.49769, recall of 0.45706 and f-measure of 0.47181. While Microsoft Word 2007 summarizer reaches the average precision 0.47242, recall of 0.40778 and f-measure of 0.43026. Baseline reaches an average precision of 0.47002, recall of 0.45624 and f-measure of 0.46108.

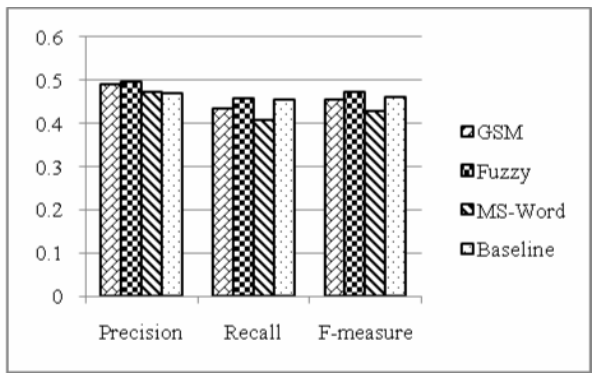

Figure 4. Average precision recall and f-measure score among four summarizers

TABLE II COMPARISON OF THE NUMBER OF DOCUMENTS FOR AVERAGE F-MEASURE SCORE FROM DIFFERENT SUMMARIZER

| Average F-measure | GSM | Fuzzy | MS-Word | Baseline |
|---|---|---|---|---|
| < 0.30000 | 8 | 1 | 9 | 5 |
| 0.30000-0.39999 | 25 | 24 | 46 | 25 |
| 0.40000-0.49999 | 51 | 54 | 38 | 54 |
| 0.50000-0.59999 | 32 | 38 | 28 | 33 |
| 0.60000-0.69999 | 8 | 7 | 4 | 7 |
| >=0.70000 | 1 | 1 | 0 | 1 |

Table II shows 32.80% of documents from GSM reaches the average f-measure more than 0.50000 while the fuzzy summarizer reaches 36.80% on the other hand 25.60% and 32.80% of Microsoft Word 2007 and baseline gets the average recall more than 0.50000.

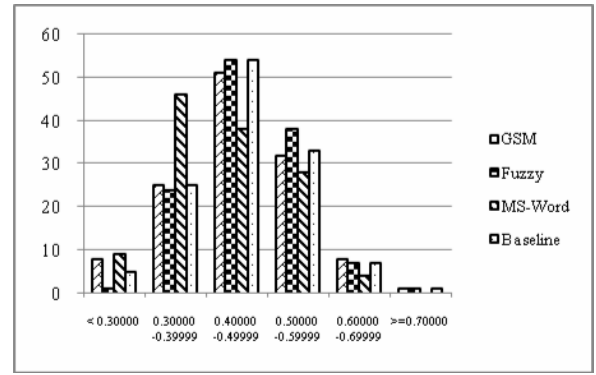

Figure 5. The number of documents for average f-measure score from different summarizer

## VI. CONCLUSION AND FUTURE WORK

In this paper, we have presented a fuzzy logic aided sentence extractive summarizer that can be as informative as the full text of a document with better information coverage. A prototype has also been constructed to evaluate this automatic text summarization scheme using as input some news articles collection provided by DUC2002. We extracted the important features for each sentence of the document represented as the vector of features consisting of the following elements: title feature, sentence length, term weight, sentence position, sentence to sentence similarity, proper noun, thematic word and numerical data.

We have done experiments with 125 data set, comparing our summarizer with Microsoft Word 2007 and baseline using precision, recall and f-measure built by ROUGE. The results show that the best average precision, recall and f-measure to summaries produced by the fuzzy method. Certainly, the experimental result is based on fuzzy logic could improve the quality of summary results that based on the general statistic method. In conclusion, we will extend the proposed method using combination of fuzzy logic and other learning methods and extract the other features could provide the sentences more important.

ACKNOWLEDGMENT

We would like to thank Suan Dusit Rajabhat University and Universiti Teknologi Malaysia for supporting us.

AUTHORS PROFILE

**Ladda Suanmali** is a Ph.D. candidate in computer science in the Faculty of Computer Science and Information Systems at Universiti Teknologi Malaysia. She graduated a bachelor degree in computer science from Suan Dusit Rajabhat University, Thailand in 1998. She graduated a master degree in information technology from King Mongkut's University of Technology Thonburi, Thailand in 2003.

Since 2003, she has been working as a lecturer in the Faculty of Science and Technology, Suan Dusit Rajabhat University. Her current research interests include text summarization, data mining, and soft computing.

**Naomie Salim** is an Associate Professor presently working as a Deputy Dean of Postgraduate Studies in the Faculty of Computer Science and Information System in Universiti Teknologi Malaysia. She graduated a bachelor degree in Computer Science from Universiti Teknologi Malaysia in 1989. She graduated a master degree in Computer Science from University of Illinois in 1992. In 2002, she received a Ph.D (Computational Informatics) from University of Sheffield, United Kingdom.

Her current research interest includes Information Retrieval, Distributed Database and Chemoinformatic.

**Mohammed Salem Binwahlan** is a Ph.D. candidate in computer science in the Faculty of Computer Science and Information Systems at Universiti Teknologi Malaysia. He received his B.Sc. degree in Computer Science from Hadhramout University of Science and Technology, Yemen in 2000. He received his Master degree from Universiti Teknologi Malaysia in 2006.

He has been working as a lecturer at Hadhramout University of Science and Technology. His current research interest includes Information Retrieval, Text Summarization and Soft Computing.